\documentclass[12pt]{iopart}
\usepackage[dvipdf]{graphicx}

\begin{document}
\newcommand{\avg}[1]{\langle{#1}\rangle}
\newcommand{\Avg}[1]{\left\langle{#1}\right\rangle}
\def\be{\begin{equation}}
\def\ee{\end{equation}}
\def\bc{\begin{center}} 
\def\ec{\end{center}}
\def\bea{\begin{eqnarray}}
\def\eea{\end{eqnarray}}
\def\bwt{\begin{widetext}}
\def\ewt{\end{widetext}}
\def\ra{\rightarrow}
\def\ba{\backslash}
\def\pa{\partial}
\title{Algorithm for counting large directed loops }

\author{Ginestra Bianconi$^1$ and Natali Gulbahce$^2$ }
\address{$^1$The Abdus Salam International Center for Theoretical 
Physics, Strada Costiera 11, 34014 Trieste, Italy\\
$^2$Theoretical Division and Center for Nonlinear Studies, Los Alamos 
National Laboratory, NM 87545, USA
}

\begin{abstract}  
We derive a Belief-Propagation algorithm for counting large  loops in a
directed network.  We evaluate the distribution of the number of small loops in
a directed random network with given degree sequence.  We apply the algorithm
to a few characteristic directed networks of various network sizes and loop
structures and compare the algorithm with exhaustive counting results when
possible. The algorithm is adequate in estimating loop counts for large
directed networks and can be used to compare the loop structure of directed
networks and their randomized counterparts. 

\end{abstract}

\maketitle
\section{Introduction}

The structure of complex networks highly affect the critical behavior
of different cooperative models \cite{Dorogovtsev} and the nonlinear
dynamical process that take place on the network \cite{Motter_rev}.
  
In particular both the directionality of the links which suggest a non
symmetric interaction \cite{Grad,Synchr,Galla} and the local loop
structure \cite{Klemm} of the network which correlates neighboring nodes has important dynamical consequences.  In fact directionality of links
becomes particularly important when a transport process of mass or information
takes place in the network \cite{Grad} and    the loop structure in these directed networks are
crucial for assessing the networks' robustness characteristics and determining
the load distribution.

Directed networks are ubiquitous in both man-made and natural systems. Some
examples of directed networks are the Texas power-grid, the World-Wide-Web, the
foodwebs and in biological networks, such as the metabolic network, the
transcription network and the neural network.
The local structure of directed network is radically different from
the structure of their undirected version \cite{last}.While many
undirected networks  are characterized but large clustering
coefficient \cite{swn} and large number of short loops \cite{Loops,loop_lungo} this is not a general trend  for directed  networks.
For example  the $C. elegans$ neural network  has a
over-representation of short loops compared to a randomized network if
the direction of the links is not considered while it has an
under-representation of the number of loops  when the
direction of the links is taken into account  \cite{last}.

Nevertheless, while counting small loops is a given network is a
relatively easy computation,  counting large loops in a real world
network is a very hard task. In fact the
number of large loops can, and usually does grow exponentially with the number
of nodes $N$ in the network. The known efficient exhaustive
algorithms \cite{Johnson,Tarjan} for counting loops still have a time bound of
$O(N*M*(L+1))$ where $N,M,L$ are respectively the number of nodes, links and
loops in the network. This task becomes computationally inapplicable for
counting large loops in many real networks. Two different approaches for the
study of long loops have been proposed: devising MonteCarlo algorithms, or
using Belief-Propagation (BP) algorithms. The two approaches have both been 
pursued in the case of undirected networks \cite{Loopy,Algorithm,Circuits}. The BP 
algorithm~\cite{Algorithm} is a heuristic algorithm which does not have sampling 
bias as the MonteCarlo algorithm~\cite{Loopy} does and is observed to give 
good results as the size of the network increases.

In this paper we generalize the BP algorithm proposed by
 \cite{Algorithm,Circuits} to directed networks. We analytically derive the
outcome of the algorithm in an ensemble of random uncorrelated networks with
given degree sequence of in/out degrees in agreement with the prediction for
the average number of nodes in this ensemble \cite{last}.  We finally study the
particular limitations of the algorithm for small network sizes and small
number of loops in the graph. The paper is divided into four further sections.
In Section 2, we derive the BP algorithm for directed networks following the
similar steps as described in \cite{Circuits}. In Section 3, we derive the
distribution of the small loops in uncorrelated random ensembles. In Sections 4
and 5, we describe the steps in the algorithm and its application to a few
characteristic directed networks.

\section{Derivation of the BP algorithm}
Given a network G of $N$ nodes and $M$ links,  
we define a partition function $Z(u)$ as the generating functions of
the number ${\cal N}_L$ of loops of length $L$ in the  network,
\be
Z(u)=\sum_Lu^L {\cal N}_L(G).
\ee 
Starting with this partition function, we can define a free energy $f(u)$ and an
entropy $\sigma(\ell)$ of the loops of length $L=N\ell$$\sigma(\ell)$ as
the following:
\bea
f(u)=\frac{1}{N}\ln Z(u)\nonumber \\
\sigma(\ell)=\frac{1}{N}\ln{\cal N}_{L=\ell N}.
\eea

For each directed link in the network, $l=\avg{ij}$ from node $i$ to node $j$,
if we define a variable $S_l = 0, 1$  which indicates if a given loop passes
through the link $l$, the partition function $Z(u)$ can then be written as 
\be
Z(u)=\sum_{\{S_l\}}w(\{S_l\})u^{\sum_{l=1}^M S_l},
\label{Z1}
\ee
where $w(\{S_l\})$ is an indicator function of the loops, i.e. it is 1 if
the variables $S_l=1$ have a support which form a closed loop, and it is zero 
otherwise. As in References~\cite{Algorithm,Circuits} we take for simplicity a relaxed local 
form  of the indicator function $w(\{S_l\})$ which is 1 also if the assignment 
of the link variables $S_l$ is compatible with a few disconnected loops.
In particular we take $w(\{S_l\})$ as 
\be
w(\{S_l\})=\prod_{i=1}^N w_i(\{S\}_i)
\ee
where $\{S\}_i=\{S_{\avg{ij}}\}_{j\in \pa i}$, and $\pa i$ indicates
the set of nodes either pointing to $i$ or pointed by $i$ and where $w_i(\{S\}_i) $ is defined as 
\bea
w_i(\{S\}_i)=\left\{\begin{array}{lcr}1&\mbox{if}\ \sum_{j\in \pa_{+}i}S_{\avg{ij}}=1
 \ \mbox{and}\  \sum_{j\in \pa_{-}i}S_{\avg{ij}}=1\nonumber\\
1&\mbox{if}\ \sum_{j\in \pa_{+}i}S_{\avg{ij}}=0\  \mbox{and}\ 
\sum_{j\in \pa_{-}i}S_{\avg{ij}}=0\nonumber \\
0&\mbox{otherwise}\end{array} \right.
\eea
with $\pa_+i$ and  $\pa_{-}i$ indicating the set of nodes $j$ which
points  to $i$  or which are pointed by $i$, respectively.
Finding the free energy $f(u)$ associated with the partition function
$(\ref{Z1})$ can be cast into finding normalized distributions
$p_v(\{S_l\})$ which minimize the Kullback distance 
\be
F_{Gibbs}[p_v]=\sum_{\{S_l\}}p_v(\{S_l\})\ln\left(\frac{p_v(\{S_l\})}{w(\{S_l\})u^{\sum_lS_l}}\right).
\ee
In fact it is straightforward to show that $F_{Gibbs}$ assumes its minimal value
when $p_v(\{S_l\})=w(\{S_l\})u^{\sum_lS_l}/Z$. If the given network is a tree, the 
trial distribution $p_v(\underline{S})$ takes the form
\be
p(\{S_l\})=\left(\prod_{l} p_l(S_l)\right)^{-1}\left(\prod_i
  p_i(\underline {S}_i)\right).
\label{tree}
\ee
with $p_{l}(S_l)$ and $p_i(\{S\}_i)$ being the marginal distributions 
\bea
p_{\ell}(S_{\ell})&=&\sum_{\{S_l\} \ba S_l}p(\{S_l\})
\nonumber \\
p_i(\{S\}_i)&=&\sum_{\{S\} \ba \{S\}_i}p(\{S\}).
\eea
In a real case, when the network is not a tree,  we can always take a variational approach and try a
given trial distribution of the form (\ref{tree}). 
After taking this variational approach, we then have to minimize the Bethe free
energy $F_{Bethe}$ as  
\be
F_{Bethe}[\{p_i\},\{p_l\}]=\sum_i \sum_{\{S\}_i\ba
  S_l}p_i(\{S\}_i)\ln\left(\frac{p_i(\{S\}_i)}{w_i(\{S\}_i)}\right)-\sum_l\sum_{S_l}p_l(S_l)\ln(p_l(S_l)u^{S_l}).
\ee
For each link $l\avg{ij}$ starting from $i$ and ending in $j$, there are the constraints
\bea
p_l(S_l)&=&\sum_{\{S\}_i}p_i(\{S\}_i)\nonumber \\
p_l(S_l)&=& \sum_{\{S\}_j}p_j(\{S\}_j).
\label{constraint}
\eea
Introducing the Lagrangian multipliers enforcing the conditions
$(\ref{constraint})$ and the normalization of the probabilities it is
easy to show that a possible parametrization of the marginals 
is the following,  
\bea
p_l(S_l)&=&\frac{1}{C_l}(u y_{i\ra j} \hat{y}_{j\ra i})^{{S}_l}\nonumber
\\
p_i(\{S\}_i)&=&\frac{1}{C_i}w_i(\{S\}_i)\prod_{j\in
  \pa_{+}i}(u y_{i\ra j})^{S_{<ij>}}\prod_{j\in
  \pa_{-}i}(u \hat{y}_{i\ra j})^{S_{<ij>}}.
\eea
For every directed link $\avg{ij}$ from node $i$ to node $j$
the values of the messages $y_{i\ra j}$ and $\hat{y}_{j \ra i}$ are
fixed by the constraints in Eq.~$(\ref{constraint})$ to satisfy the following
BP equations:
\bea
y_{i\ra j}=\frac{u\sum_{k\in \pa_{-} i} y_{k\ra
    i}}{1+u^2\sum_{k'\in\pa_{+}(i)\ba j}\hat{y}_{k'\ra i}\sum_{k\in
    \pa_{-} i} y_{k\ra i}}\nonumber \\
\hat{y}_{j\ra i}=\frac{u\sum_{k\in \pa_{+} j} \hat{y}_{k\ra
    j}}{1+u^2\sum_{k'\in\pa_{-}j\ba i}{y}_{k'\ra j}\sum_{k\in
    \pa_{+} i} \hat{y}_{k\ra j}}.
\label{BP}
\eea
The normalization constants for the marginals is consequently given by  
\bea
C_l&=&1+u y_{i\ra j}\hat{y}_{j\ra i}.\nonumber \\
C_i&=&1+u^2\sum_{k'\in\pa_{-}i}{y}_{k'\ra i}\sum_{k\in
    \pa_{+} i} \hat{y}_{k\ra i}.
\eea
The Bethe free energy density $f_{Bethe}=\frac{1}{N}F_{Bethe}$ becomes
\be
Nf_{Bethe}(u)=-\sum_{l=1}^M\ln C_l+\sum_{i=1}^N\ln C_i.
\label{f}
\ee
For any given value of $u$ the loops length is given by
\be
\ell(u)=\frac{1}{N}\sum_{l=1}^Mp_l(1)=\frac{1}{N}\sum_l\frac{u y_{i\ra
    j}\hat{y}_{j\ra i}}{1+u y_{i\ra j}\hat{y}_{j\ra i}}.
\label{ell}
\ee
The function $\ell(u)$  can be inverted giving the function $u(\ell)$
and finally proving an expression for the entropy of the loops in the
graph under a Bethe variational approach,
\be
\sigma_{Bethe}(\ell)=f(u(\ell))-\ell\ln u(\ell).
\label{sigma}
\ee
\section{Derivation of the typical number of short loops in random directed  network with given degree sequence}

We consider an ensemble of random directed networks with given
degree sequence $\{k^{in}_i,k^{out}_i\}~\forall i=1,\dots, N$.  If the
maximal in/out connectivities $K^{in}$/$K^{out}$ of the network satisfy the inequality
$K^{in}K^{out}<(\avg{k_{in}})N$, the network is uncorrelated. 
By  $q_{k_{in},k_{out}}$ we indicated the degree distribution of the
ensemble.
In Ref. \cite{last} an expression for the average number ${\cal N}_L$
of small loops was given,
\be
\avg{{\cal N}_L}\simeq \frac{1}{L}\left(\frac{\avg{k_{in}k_{out}}}{\avg{k_{in}}}\right)^L
\ee 
valid as long as 
\be
L\ll N\frac{\avg{k_{in}k_{out}}^2}{\avg{(k_{in}k_{out})^2}}.
\label{bound}
\ee
Is an interesting exercise to see what is the distribution  of the
number of  small
loops in the ensemble of directed networks by solving the BP equation
for a random directed ensemble in parallel with the distribution found
in the undirected case  \cite{Circuits}. 
In a directed network ensemble the  BP  messages $y$
and $\hat{y}$ along each link are  equally distributed depending only on the
value of $u$.  Given the BP equations $(\ref{BP})$, the distribution $P(y;u)$ of the
field $y$ has to satisfy the self-consistent equation 
\bea
P(y;u)&=&
\sum_{k_{out}=1}^{\infty}\frac{k_{out}}{\avg{k_{out}}}q_{0,k_{out}}\delta(y)+\sum_{k_{in}=1}^{\infty}
\sum_{k_{out}=1}^{\infty}\frac{k_{out}}{\avg{k_{out}}}q_{k_{in},k_{out}}\nonumber \\
& &\int_{0}^{\infty} dy_1 P(y_1;u) \dots, dy_{k_{in}}P(y_{k_{in}};u)\nonumber \\
& & \int_0^{\infty}d\hat{y}_1 P(\hat{y}_1;u) \dots
d\hat{y}_{k_{out}}P(\hat{y}_{k_{out}};u)\delta(y-g_k(\{y\},\{\hat{y}\}))
\label{uno}
\eea
with 
\bea
g_1=u y_1\nonumber \\
g_k=\frac{u\sum_{k\in \pa_{-} i} y_{k\ra
    i}}{1+u^2\sum_{k'\in\pa_{+}(i)\ba j}\hat{y}_{k'\ra i}\sum_{k\in
    \pa_{-} i} y_{k\ra i}} \ \mbox{for}\ k\geq2.
\eea
In fact, given a random edge the probability that its starting node
$i$ has connectivity $(k_{out},k_{in})$ is given by
$\frac{k_{out}}{\avg{k_{out}}}q_{k_{in},k_{out}}$. 
The fields $\hat{y}$ have to satisfy a similar recursive equation, i.e.
\bea
P(\hat{y};u)&=& \sum_{k_{in}=1}^{\infty}
\frac{k_{in}}{\avg{k_{in}}}q_{k_{in},0}\delta(y)+\sum_{k_{in}=1}^{\infty}
\sum_{k_{out}=1}^{\infty}\frac{k_{in}}{\avg{k_{in}}}q_{k_{in},k_{out}}\nonumber \\
& &\int_{0}^{\infty} dy_1 P(y_1;u) \dots
du_{k_{in}}y_{k_{in}}P(y_{k_{in}};u)\nonumber \\
& &\int_0^{\infty}d\hat{y}_1 P(\hat{y}_1;u) \dots
du_{k_{out}}\hat{y}_{k_{out}}P(\hat{y}_{k_{out}+};u)\delta(y-\hat{g}_k(\{y\},\{\hat{y}\}))
\label{due}
\eea
with 
\bea
\hat{g}_1=u \hat{y}_1\nonumber \\
\hat{g}_k=\frac{u\sum_{k\in \pa_{+} i} \hat{y}_{k\ra
    i}}{1+u^2\sum_{k'\in\pa_{+}(i)\ba j}\hat{y}_{k'\ra i}\sum_{k\in
    \pa_{-} i} y_{k\ra i}} \ \mbox{for}\ k\geq2.
\eea
For a given small value of $ u=u_m+\epsilon$, the two coupled equations in 
Eq.~$(\ref{uno})$ and Eq.~$(\ref{due})$ become independent. 
By proceeding as in \cite{Circuits}, we find that the number of small
loops in the ensemble is given by 
\be
\avg{N_L}\simeq\frac{1}{L}\left(\frac{\avg{k_{in}k_{out}}}{\avg{k_{in}}}\right)^L
\label{short}
\ee
with Poisson fluctuations for loops of size $L\ll\log(N)$. For larger
loop sizes up to the boundary limit given by $(\ref{bound})$, the average number of loops in the ensemble is still given
by $(\ref{short})$ but with significant fluctuations in the number of loops.

\section{The BP algorithm}

The study of the partition function Eq. $(\ref{Z1})$ carried on in
Section 2 is such that a new algorithm for counting large loops in a
directed network can be formulated. In particular,
given a network with $N$ nodes and $M$ links, the algorithm is: 

\begin{itemize}
\item  
Initialize the messages $y_{i\ra j}$, $\hat{y}_{j\ra i}$ for every
directed link between $i$ and $j$ to random values.
\item
For every value of $u$, iterate the BP equations in Eq.~$(\ref{BP})$ 
\bea
y_{i\ra j}=\frac{u\sum_{k\in \pa_{-} i} y_{k\ra
    i}}{1+u^2\sum_{k'\in\pa_{+}(i)\ba j}\hat{y}_{k'\ra i}\sum_{k\in
    \pa_{-} i} y_{k\ra i}}\nonumber \\
\hat{y}_{j\ra i}=\frac{u\sum_{k\in \pa_{+} j} \hat{y}_{k\ra
    j}}{1+u^2\sum_{k'\in\pa_{-}j\ba i}{y}_{k'\ra j}\sum_{k\in
    \pa_{+} i} \hat{y}_{k\ra j}}.
\eea
until convergence.
\item
Calculate $\ell(u)$ and $f(u)$ from Eqn's $(\ref{ell})$ and $(\ref{f})$
which we recall here for convenience
\be
\ell(u)=\frac{1}{N}\sum_{l=1}^Mp_l(1)=\frac{1}{N}\sum_l\frac{u y_{i\ra
    j}\hat{y}_{j\ra i}}{1+u y_{i\ra j}\hat{y}_{j\ra i}}.
\ee
\bea
Nf_{Bethe}(u)&=&-\sum_{l=1}^M\ln \left(1+u y_{i\ra j}\hat{y}_{j\ra
  i}\right)\nonumber \\
&& +\sum_{i=1}^N\ln \left(1+u^2\sum_{k'\in\pa_{-}i}{y}_{k'\ra i}\sum_{k\in
    \pa_{+} i} \hat{y}_{k\ra i}\right).
\eea
\item
Evaluate $\sigma(\ell)$ by Eq. $(\ref{sigma})$ which again we repeat
here for convenience 
\be
\sigma_{Bethe}(\ell(u))=f(u)-\ell(u)\ln u.
\label{sigma}
\ee
\end{itemize}

\begin{figure}
 \includegraphics[width=6.5cm, height=5.5cm]{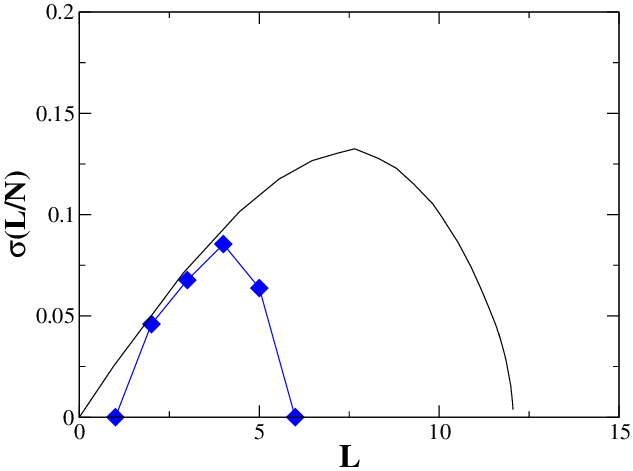}
 \caption{Entropy $\sigma(L/N)$ of the loops of length $L$ for the
   real {\it Chesapeake} food-web  (solid line) and the entropy of the
   loops counted  by exact enumeration  (diamods). }
 \label{chesapeake}
\end{figure}

\section{Application of the algorithm to real directed networks}

We applied the formulated algorithm to a large set of directed networks
 \cite{nnets}. For some of these networks we calculated the number of
 loops ${\cal N}_L$ of lenght $L$ directly by exact enumeration
 \cite{Tarjan}. 
We then compare the entropy of the loops
$\sigma(\ell)$ find by the BP algorithm with
the entropy of the loops $\sigma_0(\ell)$ find by directed enumeration
of the number of loops
\be
\sigma_0(\ell)=\frac{1}{N}\ln({\cal N}_{L=\ell N}^{\mbox{exact}})
\ee
 We note that for the foodweb with
small number of nodes the algorithm does not provide a good approximation for
the number of loops present in the graph. A dramatic example is the
Chesapeake foodweb. In this case we were able to count all the loops in the
network exhaustively since the network contains very few loops. 
 In this case the BP algorithm since the loops are few the BP
 algorithm highly overestimates the largest loop in the
network (See Figure \ref{chesapeake}). In fact it predict a largest loop of lend $L_{max}=12$ where
the largest loop is of length $L_{max}=7$. This effect is observed to be present also in the undirected BP algorithm
 \cite{Algorithm}. 

\begin{figure}
 \includegraphics[width=7.5cm, height=5.5cm]{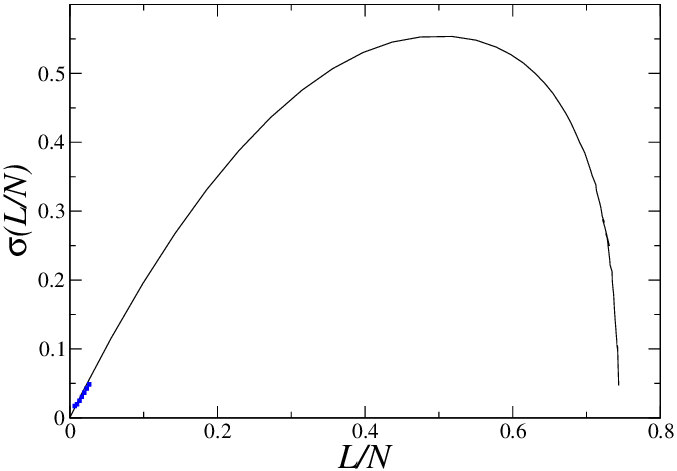}
 \caption{Entropy $\sigma(L/N)$ of the loops of length $L$ for the
   real {\it C.elegans} neural network (solid line) and the entropy of the loops counted  by
   exact enumeration for small loops (small diamods). \\}
 \label{celegans}
\end{figure}
The discrepancy is predicted to be strong only in cases where the 
size of the network is small  and the number of loops in the network is
small just as in the Chesapeake case.  When the network has a larger number of
loops and the entropy of the loops is larger, much better results are expected.
In the case of the {\it C. elegans} neural network ($N=306$) the entropy for
small number of loops is overlapping with the results  of exact enumeration as 
it can clearly be seen in Figure \ref{celegans}.
We further compare the results of the algorithm on a given network and
on randomized network ensemble.
A typical example is the metabolic network of {\it E. coli}
 \cite{nnets} in which we could compare the entropy  provided by the BP
algorithm with the entropy of a series of $100$ random network with
the same degree distribution.

\section{Conclusions}
In conclusion we provide a new algorithm for counting large loops in
directed network.
The algorithm is predicted to give good results only for large
networks size $N$. In the paper we demonstrate cases in which it fails
to predict the right entropy and loop structure due to the small size of the
network. We propose to study the significance of loops structure in
large networks by comparing the results of the algorithm on real
networks and randomized networks when networks are large an the number
of loops in the network are also large.

\section{Acknowledgment}
We acknowledge G. Semerjian and A. E. Motter for interesting discussions.
\\

 \begin{figure}
 \includegraphics[width=7.5cm, height=5.5cm]{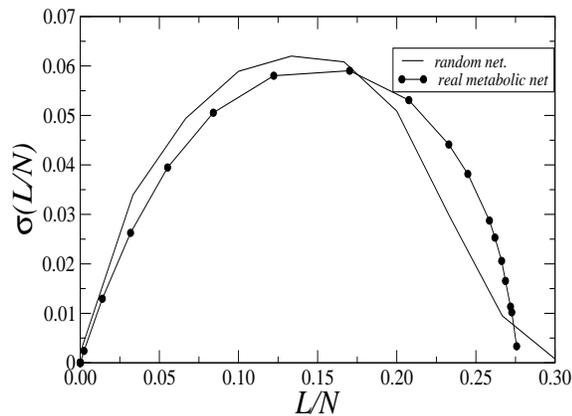}
 \caption{Entropy $\sigma(L/N)$ of the loops of length $L$ for the
   real metabolic network and average entropy of the loops in the
   randomized network ensemble with same degree sequence. }
 \label{metabolic}
\end{figure}

\end{document}